\baselineskip=18pt
\def\a{\alpha}\def\c{\chi}\def\d{\delta}
\def\f{\phi}
\def\l{\lambda}\def\m{\mu}\def\n{\nu}\def
\p{\pi}\def\r{\rho}\def\s{\sigma}
\def\x{\xi}\def\z{\zeta}

\def\O{\Omega}

\def\de{\partial}
\def\inf{\infty}\def\id{\equiv}\def\mo{{-1}}

\def\const{{\rm const}}

\def\exp#1{{\rm e}^{#1}}

\def\af{asymptotically flat }\def\hd{higher derivative }
\def\fe{field equations }\def\bh{black hole }

\def\ssy{spherically symmetric }

\def\hdim{higher dimensional }
\def\sch{Schwarzschild }

\def\GB{Gauss-Bonnet }

\def\dsy{dynamical system }
\def\ab{asymptotic behavior}

\def\dsy{dynamical systems }

\def\section#1{\bigskip\noindent{\bf#1}\smallskip}
\def\nota{\footnote{$^\dagger$}}

\def\PL#1{Phys.\ Lett.\ {\bf#1}}
\def\PRL#1{Phys.\ Rev.\ Lett.\ {\bf#1}}
\def\PR#1{Phys.\ Rev.\ {\bf#1}}\def\CQG#1{Class.\ Quantum Grav.\ {\bf#1}}
\def\NP#1{Nucl.\ Phys.\ {\bf#1}}
\def\JMP#1{J.\ Math.\ Phys.\ {\bf#1}}

\def\MPL#1{Mod.\ Phys.\ Lett.\ {\bf #1}} 
\def\PRep#1{Phys.\ Rep.\ {\bf#1}}
\def\AoM#1{Ann.\ Math.\ {\bf#1}}

\def\ref#1{\medskip\everypar={\hangindent 2\parindent}#1}
\def\beginref{\begingroup
\bigskip
\centerline{\bf References}
\nobreak\noindent}
\def\endref{\par\endgroup}

\def\er{{\cal R}}\def\es{{\cal S}}\def\ef{e^{-2\f}}
\def\ter{{1\over3}}\def\tV{\tilde V}


{\nopagenumbers
\vskip80pt
\centerline{\bf Global properties of dilatonic Gauss-Bonnet black
holes}
\vskip40pt
\centerline{{\bf M. Melis}\footnote{$^\dagger$}{e-mail:
maurizio.melis@ca.infn.it} and
{\bf S. Mignemi}\footnote{$^\ddagger$}{e-mail:
smignemi@unica.it}\footnote{$^*$}{and INFN, Sezione di Cagliari}}
\vskip10pt
\centerline {Dipartimento di Matematica, Universit\`a di Cagliari}
\centerline{viale Merello 92, 09123 Cagliari, Italy}

\vskip100pt
\centerline{\bf Abstract}
\vskip10pt
{\noindent
We study the phase space of the \ssy solutions of Einstein-\GB system
nonminimally coupled to a scalar field and show that in four dimensions
the only regular \bh solutions are asymptotically flat.}
\vskip100pt\
P.A.C.S. Numbers: 97.60.Lf 11.25.Mj
\vfil\eject}

\section{1. Introduction}

In the context of quantum gravity, it has been often advocated
the possibility that higher-derivative corrections should be added
to the Einstein-Hilbert (EH) lagrangian of general relativity, in order
to obtain a better behaved theory. Among the various
possibilities, an especially interesting class is given by
the so-called Lovelock Lagrangians.

These were introduced in [1] as the only possible generalization
to higher dimensions of the EH lagrangian that gives rise to \fe
which are second order in the metric, linear in the second derivatives,
and divergence free.

Each term $L_i$ of the Lovelock Lagrangian consists in a combination of
contractions of a given power of the Riemann tensor, the lowest order
terms $L_0$ and $L_1$ being the cosmological constant and the EH
Lagrangian, respectively. The $i$-th term is defined in dimensions $d\ge2i$,
and in $d=2i$ reduces to a total derivative, that takes the form of
the \GB topological invariant. For this reason, the Lovelock Lagrangians
are sometimes called \GB (GB) lagrangians. In the following, for simplicity
of notation, we reserve this name for the quadratic term
$$L_2=\sqrt{-g}\,(\er^{\m\n\r\s}\er_{\m\n\r\s}-4\er^{\m\n}\er_{\m\n}+\er^2),$$
which gives the lowest order correction to the EH lagrangian and hence is
the most common in applications. This quadratic combination was originally
introduced by Lanczos [2].

From the topological origin of the Lovelock terms it also follows
the important property that they do not introduce any new propagating
degrees of freedom in the spectrum [3]. In particular, $L_2$ is the only
quadratic combination of curvature terms that does not give rise to tachyons
or ghosts.
In four dimensions it reduces to a total derivative, and hence does
not contribute to the field equations unless nonminimally coupled to a scalar
field.

Because of their properties, the Lovelock Lagrangians were
widely investigated in the context of Kaluza-Klein theories [4]. Later it was
realized that they intervene also in the low-energy limit of string
theories, where the GB term, nonminimally coupled to the dilaton, gives a
contribution to the effective action [3,5].

The presence of higher-order terms in the action may smooth the behaviour
of the solutions in the regime of strong gravity, as for example near the
initial singularity of cosmology, and can induce spontaneous compactification.
Cosmological solutions of EH-GB models were therefore investigated in [6].
More recently, a renewed interest in \GB cosmology has arisen in the context
of the braneworld scenario [7].

Lovelock Lagrangians are also of interest for black hole physics.
In fact, in addition to the modification of the short-distance behaviour of
the solutions near the singularities, also global properties may be altered
if one allows for the possibility of nonminimal coupling of scalars as
required by the string effective action, since the standard no-hair theorems
[8] can be circumvented in that case.

Exact \bh solutions of the higher-dimensional EH-GB model in the
absence of scalar fields were obtained in [9], and exhibit either Minkowski
or (anti-)de Sitter asymptotics. The more difficult case
of the EH-GB theory coupled to a scalar field in four dimensions
was originally studied perturbatively in [10], using the methods
introduced in [11] for higher dimensions, and then
numerically in [12]. It turned out that regular \af \bh solutions
exist with nontrivial scalar field. The scalar charge is however a
function of the mass (and charge) of the black hole\nota{An
independent scalar charge is however present in
the case of multiple scalar fields [13].}.

In these papers the existence of \bh solutions that are not \af
was not investigated. However, it is known that in $D>4$ the EH-GB
theory with no scalar possesses (anti-)de Sitter solutions [3-4].
Although in [14] it was argued that no asymptotically (anti-)de
Sitter solutions can be present if a nonminimally coupled scalar
field is added to the model, it is interesting to seek for more
general solutions of the EH-GB system, with arbitrary asymptotic
behaviour.

A powerful technique for investigating this topic is the study of
the phase space of the solutions of the field equations. This
method has been used in several models of Einstein-Maxwell theory
nonminimally coupled to scalars [15]. As mentioned above, when a GB term
is added to the action, the field equations are still second order,
and linear in the second derivatives, but no longer quadratic
in the first derivatives.
This fact gives rise to several technical problems. In
particular, the potential of the dynamical system is no longer
polynomial, but presents poles for some values of the variables.

In this paper, we use the methods of the theory of dynamical systems
to investigate the case of EH-GB theory coupled to scalars.
The interest of this study relies
both in the possibility of classifying EH-GB black holes and in the
demonstration of the applicability of these methods also to
higher-derivative models of the Lovelock type.

The result of our investigation is that in four dimensions only \af
regular \bh solutions can exist, while for $D>4$ we are not able to
exclude the possibility of more general asymptotics.

\section{2. The dynamical system for 4-dimensional spacetime}
We start by considering the four-dimensional case and will then
generalize to higher dimensions. Since in four dimensions the GB
term is a total derivative in absence of scalars, the calculations
are simpler, and the results may be qualitatively different from
those obtained in higher dimensions.

The four-dimensional action is
$$I={1\over16\p G}\int\sqrt{-g}\ d^4x
\left[\er-2(\de\f)^2+\a\ef\es\right],\eqno(2.1)$$
where $\es=\er^{\m\n\r\s}\er_{\m\n\r\s}-4\er^{\m\n}\er_{\m\n}+\er^2$
is the GB term, $\f$ is a scalar field and $\a$ is a constant with
dimension (length)$^2$. This action also emerges as a low-energy
expansion of string theory [3,5]. In this case the constant $\a$ is
proportional to the slope parameter $\a'$.

We adopt the \ssy ansatz [15]
$$ds^2=-e^{2\n}dt^2+\l^{-2}e^{2(2\z-\n)}d\x^2+e^{2(\z-\n)}d\O_2^2,
\qquad\f=\n-\c,\eqno(2.2)$$ where $\n$, $\z$, $\c$ and $\l$ are
functions of $\x$ and $d\O_2^2$ is the volume element of a
2-sphere.

Substituting the ansatz (2.2) into (2.1) and performing some
integrations by parts, the action can be cast in the form
$$\eqalignno{
I=\, -2\int d^4x\bigg\{ &\l\Big[2\n'^2-\z'^2+\c'^2
-2\n'\c'+8\a\n'(\n'-\c')e^{2\c-2\z}\Big]&\cr
&-8\a\l^3\n'(\n'-\c')(\z'-\n')^2e^{2\c-4\z}-{1\over\l}e^{2\z}\bigg\}.&(2.3)}
$$
As usual, the action (2.3) does not contain derivatives of $\l$, which
acts therefore as a Lagrangian multiplier enforcing the Hamiltonian
constraint.
Another relevant property of the action is that, in spite of the
presence of the \hd GB term, after integration by parts it contains only
first derivatives of the fields, although up to the fourth power, and
therefore gives rise to second order field equations.

We vary the action and then write the \fe in first order form in terms of
the new variables,
$$X=\z',\quad Y=\n',\quad W=\c',\quad Z=e^{\z},\quad V=\sqrt\a\,e^{\c},$$
which satisfy
$$Z'=XZ,\qquad V'=WV.\eqno(2.4)$$

Varying with respect to $\l$ and then choosing the gauge $\l=1$, one obtains
the Hamiltonian constraint
$$
E\id2Y^2-X^2+W^2-2YW+Z^2+8Y(Y-W)\Big[Z^2-3(Y-X)^2\Big]{V^2\over Z^4}=0,\eqno(2.5)
$$
while variation with respect to $\n$, $\z$ and $\c$ gives the other
\fe
$$\eqalignno{
&2Y'-W'+4\left\{\Big[(2Y-W)Z^2-(Y-X)(4Y^2-2XY+XW-3YW)\Big]{V^2\over Z^4}
\right\}'=0,&(2.6)\cr
&X'-8\left\{Y(Y-W)(Y-X){V^2\over Z^4}\right\}'=Z^2+8Y(Y-W)\Big[Z^2-2(Y-X)^2\Big]
{V^2\over Z^4},&(2.7)\cr
&Y'-W'+4\left\{Y\Big[Z^2-(Y-X)^2\Big]{V^2\over Z^4}\right\}'=
-8Y(Y-W)\Big[Z^2-(Y-X)^2\Big]{V^2\over Z^4}.&(2.8)}
$$
In these variables, the problem takes the form of a five-dimensional
dynamical system, subject to a constraint. Notice that the function $E$
defined in (2.5) is a constant of the motion of the system, whose value is
set to zero by the Hamiltonian constraint.

When the coefficient of $V^2/Z^4$ in (2.5) does not vanish, one can solve
for $V^2/Z^4$ in terms of the other variables as
$${V^2\over Z^4}=-{P^2\over8Y(Y-W)[Z^2-3(Y-X)^2]},
\eqno(2.9)$$
where
$$P^2=2Y^2-X^2+W^2-2YW+Z^2,\eqno(2.10)$$
and then substitute in the other field equations.
In this way one obtains a four-dimensional system, with equations
$$\eqalignno{
&2Y'-W'-\left[{(2Y-W)Z^2-(Y-X)(4Y^2-2XY-3YW+XW)\over2Y(Y-W)[Z^2-3(Y-X)^2]}P^2
\right]'=0,&(2.11)\cr
&&\cr
&X'+\left[{(Y-X)\over Z^2-3(Y-X)^2}P^2\right]'=Z^2-{Z^2-2(Y-X)^2\over
Z^2-3(Y-X)^2}P^2,&(2.12)\cr
&&\cr
&Y'-W'-\left[{Z^2-(Y-X)^2\over2(Y-W)[Z^2-3(Y-X)^2]}P^2\right]'=
{Z^2-(Y-X)^2\over Z^2-3(Y-X)^2}P^2,&(2.13)\cr
&&\cr
&Z'=XZ.&(2.14)}$$
The substitution (2.9) allows one to avoid the singularity of the
differential equations at $Z=0$,
but introduces new singularities at the poles of (2.9), that
will complicate the discussion of the phase space.

The system may be put in canonical form solving
(2.11)-(2.13) for the first derivatives, as
$$X'=F(X,Y,W,Z),\qquad Y'=G(X,Y,W,Z),\qquad W'=H(X,Y,W,Z).\eqno(2.15)$$
The expressions so obtained are awkward and shall not be reported
here.

Another possible simplification of the system arises from the fact
that, in analogy with
the EH theory, the action depends on $\n$ only through its first
derivatives, and therefore its variation (2.6)  with respect to $\n$
determines a first integral, that can be written as:
$$2Y-W-{(2Y-W)Z^2-(Y-X)(4Y^2-2XY-3YW+XW)\over 2Y(Y-W)[Z^2-3(Y-X)^2]}P^2=a.
\eqno(2.16)$$
One may exploit this fact to reduce the problem to
the study of a three-dimensional dynamical system, by eliminating
another dynamical variable, say $W$, by means of (2.16). However,
(2.16) gives rise to a complicated third order algebraic equation,
that may have more than one solution for given values of $X$, $Y$
and $Z$. For this reason we prefer to consider the phase space as
a three-dimensional surface embedded in a four-dimensional space.

In the discussion of the dynamical system, the knowledge of the exact
solution for $\a=0$, corresponding to $V\id0$, will prove useful.
In fact for $\a=0$ the GB term is decoupled and hence, by the
no-hair theorems, one should recover the usual \sch solution.
In this case, the \fe reduce to
$$\z''=e^{2\z},\qquad\n''=0,\qquad \c''=0.$$
Their integration yields
$$e^\z={2ae^{a\x}\over1-e^{2a\x}},\qquad e^\n=e^{b\x},
\qquad e^\c=e^{c\x},$$
with $a$, $b$, $c$ integration constants.
Substituting in the constraint (2.5), one gets $a^2-2b^2+2bc-c^2=0$.
If one also requires that the solution has a regular horizon, one
must impose $a=b$, which, combined with the previous constraint,
yields $a=b=c$.
Substituting these results in the metric (2.2) and defining a new
coordinate $r=2a/(1-e^{2a\x})$, one obtains the \sch metric with
mass $a$ and constant dilaton:
$$ds^2=-(1-2a/r)dt^2+(1-2a/r)^\mo dr^2+r^2d\O^2,\qquad\f=\const.$$
In the variables of the dynamical system the solution reads
$$X=a\coth a\x,\quad Y=a,\quad W=a,
\quad Z={a\over\sinh a\x}.$$
For $a\to0$, the solution reduces to flat space.

\section{3. The phase space of the 4-dimensional solutions}
The discussion of the phase space is complicated by the presence
in the differential equations (2.11)-(2.13) of a denominator that
diverges for $Y=0$, $Y=W$, or $Z^2=3(Y-X)^2$, corresponding to infinite
or undetermined values of $V$. The hypersurfaces defined by these
conditions
are therefore singular for the trajectories. Although the solutions
of interest do not lie on these hypersurfaces, their endpoints may.
One must therefore be careful in the discussion of critical points.
It must also be noticed that the $Z<0$ region is simply a copy of
the $Z>0$ region, and hence we shall not discuss it.

The critical points of the dynamical system are defined as the points
where the derivatives of all the variables vanish. Setting to zero the
r.h.s. of (2.12)-(2.14), one obtains the following solutions:
\medskip
I) $Z_0=0$, $P_0=0$.

{\noindent The constraint (2.16) then implies $W_0=2Y_0-a$ and
$X_0=\pm\sqrt{2Y_0^2-2aY_0+a^2}$. From (2.9), it follows that $V_0=0$.
This set of critical points lies on a hyperbola in the
$X$-$Y$ plane and on a straight line in the $W$-$Y$ plane.}
\medskip
II) $X_0=0$, $Y_0=W_0=\pm Z_0$.

{\noindent These points are placed on the singular hypersurface
$Y=W$, and in fact $V\to\inf$ there, except when $Y_0=W_0=Z_0=0$.
Therefore, only this case must be considered. It coincides
with the limit $a\to0$ of case I.}

\medskip
The trajectories of the dynamical system describe the behavior of the
solutions between two critical points.
Critical points at finite distance in phase space can correspond either
to singularities or to horizons of the physical metric.
At a horizon, $e^{2\n}$ vanishes, while the radius $R\id e^{\z-\n}$
of the 2-sphere takes a finite value. Instead, a critical point with
$R=0$ corresponds in general to a singularity.
Moreover, we require that also the dilaton is regular at the horizon.
This happens if $e^{-2\f}\id e^{2(\c-\n)}$ is finite there.

These conditions are equivalent to the requirement that at the critical
point $Z^2\sim U^2\sim V^2$, where $U^2=e^{2\n}$.
The critical points at finite distance are attained for $\x\to-\inf$,
if $X_0,Y_0,W_0>0$, or for $\x\to+\inf$, if $X_0,Y_0,W_0<0$, and
the metric functions behave there as
$Z^2\sim e^{2X_0\x}$, $U^2\sim e^{2Y_0\x}$, $V^2\sim e^{2W_0\x}$.
It follows that the critical points  describe a regular horizon rather
than a naked singularity iff $X_0=Y_0=W_0=a$.
It must be noticed that these points lie on the singular hypersurfaces,
so one must be careful in taking the limits when approaching them.

The behaviour of the solutions near the critical points can be
studied by linearizing the field equations.
Unfortunately, as already noticed, in this
limit (2.9) becomes undetermined and therefore the phase space is
effectively five-dimensional near these points. As a consequence,
the linearized equations depend on how the limit $X=W=Y$ is taken.
A more rigorous treatment must be performed in the full five-dimensional
phase space, and will be exposed in more detail in the appendix.

From this investigation it results that for $a>0$ the critical points
act as repellors, while for $a<0$ they attract the trajectories.
In the limit $a=0$ all eigenvalues are null, and correspond in fact to
constant solutions.

For the investigation of the \ab of the solutions it is useful to
study the critical points at infinity [15].
In fact, spatial infinity is achieved for $R\id e^{\z-\n}\to\inf$,
i.e. $\exp\z=Z\to\inf$. This limit, which in general is reached for a
finite value $\x_0$ of $\x$, corresponds to the surface at infinity of
phase space, where the fields diverge.

Phase space at infinity may be studied by defining new variables
$$u={1\over X},\quad y={Y\over X},\quad w={W\over X},\quad z={Z\over X},
\quad p={P\over X}.\eqno(3.1)$$
The field equations at infinity are then obtained for $u\to0$ and may
be written as
$$\dot u=-fu,\qquad\dot y=g-fy,\qquad\dot w=h-fw,\qquad\dot z=(1-f)z,
\eqno(3.2)$$
where a dot denotes $u\,d/d\x$ and $f=u^2F(1,y,w,z)$, $g=u^2G(1,y,w,z)$,
$h=u^2H(1,y,w,z)$. The constraint (2.16) reduces to
$$2y-w+{(2y-w)z^2-(y-1)(4y^2-3yw-2y+w)\over2y(y-w)[z^2-3(y-1)^2]}\,p^2=0.
\eqno(3.3)$$

The critical points at infinity correspond to the vanishing of the
derivatives in (3.2) for $u\to0$.
This occurs for
\medskip
I) $y_0=w_0=0$, $z_0=\pm1$, i.e.\ $X_0=\pm\inf$, $Y_0=W_0=0$, $Z_0=\pm X_0$.
In this case $v_0$ is undetermined.
These are also the endpoints of the trajectories lying in the $V=0$
hypersurface, corresponding to the exact solutions discussed at the
end of the previous section.
\medskip
II) $z_0=0$, $p_0=0$, that in view of (3.3) yields $y_0={1\over\sqrt2}$,
$w_0=\sqrt2$. These points correspond to $X_0=\pm\inf$,
$Y_0={X_0\over\sqrt2}$, $W_0=\sqrt2\,X_0$, $Z_0=0$. It follows that $v_0=0$.
These are the endpoints of the hyperboloid $Z^2=P^2=0$.
\medskip
Also in this case, phase space is five-dimensional near the critical
points, and its linearization will be studied in more detail in the
appendix.
It will result that for $X_0>0$, points I attract all the trajectories,
while points II repel the trajectories at infinity, and are centers
for those coming from finite distance. For $X_0<0$ repellors become
attractors and viceversa.

From the location of the critical points at infinity, one can
deduce the asymptotic behaviour of the solutions [15]. In case I,
$\x\to\x_0$, and
$$e^\z\sim|\x-\x_0|^\mo,\quad e^\n\sim\const,\quad e^\c\sim\const,$$
and therefore, defining $r\sim|\x-\x_0|^\mo$ and recalling (2.2), one
obtains after some calculations the \ab
$$ds^2\sim-dt^2+dr^2+r^2d\O_2^2,\qquad\f\sim\const.$$
Hence these points correspond to \af solutions with asymptotically
constant scalar field.

At points II, $\x\to\pm\inf$, depending on $X_0$ being $\pm\inf$, and
$$e^\z\sim e^{\mp\x},\quad e^\n\sim e^{\mp\x/\sqrt2},
\quad e^\c\sim e^{\mp\sqrt2\,\x},$$
and therefore, defining $r\sim e^{\mp2\x}$,
$$ds^2\sim-r^{1/\sqrt2}dt^2+r^{-1/\sqrt2}dr^2+r^{1-1/\sqrt2}d\O_2^2,
\qquad\ef\sim r^{1/\sqrt2}.$$

Summarizing, from the study of phase space emerges that all regular
\ssy \bh solutions start from the point $X_0=Y_0=W_0=a>0$,
$Z_0=0$ and end at the critical point I (with $X>0$) at infinity
(or start at the critical point I at infinity with $X<0$ and end at
$Z_0=0$, $X_0=Y_0=W_0=a<0$).
It follows that all solutions are asymptotically flat, with asymptotically
constant scalar field. In particular, no solutions with anti-de Sitter
asymptotics exist, in accordance with the results of [14].
\bigbreak

\section{4. The dynamical system in $(m+2)$-dimensional spacetime}
In this section we generalize the previous results to the case of
$m+2$ dimensions. Since the procedure is identical to the case $m=2$,
we only give the main formulas.

The ($m+2$)-dimensional action can be written as
$$
I=\int\sqrt{-g}\
d^{(m+2)}x\left[\er-m(m-1)(\de\f)^2+\a\ef\es\right].
\eqno(4.1)$$
We adopt the \ssy ansatz
$$ds^2=-e^{2(m-1)\n}dt^2+\l^{-2}e^{2(m\z-\n)}d\x^2+e^{2(\z-\n)}d\O_m^2,
\qquad\f=\n-\c.\eqno(4.2)$$
Substituting in the action and after integration by parts, one has
$$\eqalignno{
I=\ m(m-1)&\int d^{(m+2)}x\ \bigg\{
{1\over\l}\Big[e^{2(m-1)\z}+(m-2)(m-3)\a\,e^{2\c+2(m-2)\z}\Big]\cr
&-\l\Big[2\n'^2-\z'^2+\c'^2-2\n'\c'\Big]
-2\a\,\l\, e^{2\c-2\z}\Big[(m^2-m+2)\n'^2\cr&-(m-2)(m-3)\z'^2
-4\n'\c'-4(m-2)\z'\c'\Big]\cr&+{\a\over3}\,\l^3\,e^{2\c-2m\z}(\z'-\n')^2
\Big[3(m^2+3m-2)\n'^2
-(m-2)(m-3)\z'^2\cr&-2(m-2)(m-3)\z'\n'-8(2m-1)\n'\c'-8(m-2)\z'\c'\Big]
\bigg\}.&(4.3)}
$$
As for $m=2$, the higher-dimensional action (4.3) has the property
of containing
only first derivatives of $\n$, $\z$ and $\c$ and no derivative of $\l$.
However, more terms appear in the action, because the GB term is no
longer a total derivative in absence of scalar fields.

We vary the action and write the \fe in terms of the new variables,
$$X=\z',\quad Y=\n',\quad W=\c',\quad Z=e^{(m-1)\z},\quad V=\sqrt\a\,e^\c,$$
which satisfy
$$Z'=(m-1)XZ,\qquad V'=WV.\eqno(4.4)$$

Varying with respect to $\l$ and then putting $\l=1$, one obtains
the Hamiltonian constraint
$$\eqalignno{
&P^2+\bigg\{(m-2)(m-3)Z^4
+2\Big[\,(m^2-m+2)Y^2-(m-2)(m-3)X^2-4YW\cr
&-4(m-2)XW\Big] Z^2-(X-Y)^2\Big[\,3(m^2+3m-2)Y^2
-(m-2)(m-3)X^2\cr
&-2(m-2)(m-3)XY-8(2m-1)YW-8(m-2)XW\Big]\bigg\}
{V^2\over Z^{2m/(m-1)}}=0,&(4.5)}
$$
where $P^2$ is given by (2.10).

Variation with respect to $\n$, $\z$, $\c$, respectively, gives the other
\fe
$$\eqalignno{
&2Y'-W'+2\bigg\{\bigg[\Big((m^2-m+2)Y-2W\Big)Z^2+(X-Y)\Big((m^2+3m-2)Y^2\cr
&-(m^2-m+2)XY-2(2m-1)YW+2XW\Big)\bigg]{V^2\over Z^{2m/(m-1)}}\bigg\}'=0,&(4.6)\cr
&\cr
&X'+2\bigg\{\bigg[(m-2)\Big((m-3)X+2W\Big) Z^2+\ter(X-Y)\Big(\,2m(m+1)Y^2-(m-2)(m-3)X^2\cr
&-(m-2)(m-3)XY-6mYW-6(m-2)XW\Big)\bigg]{V^2\over Z^{2m/(m-1)}}\bigg\}'=\cr
&(m-1)Z^2+\bigg[(m-2)^2(m-3)Z^4+2\Big((m^2-m+2)Y^2-(m-2)(m-3)X^2-4YW\cr
&-4(m-2)XW\Big)Z^2-{m\over3}(X-Y)^2\Big(\,3(m^2+3m-2)Y^2-(m-2)(m-3)X^2\cr
&-2(m-2)(m-3)XY-8(2m-1)YW-8(m-2)XW\Big)\bigg]{V^2\over Z^{2m/(m-1)}},&(4.7)\cr
&\cr
&Y'-W'+4\bigg\{\bigg[\Big(\,Y+(m-2)X\Big)Z^2-\ter(X-Y)^2\Big((2m-1)Y+(m-2)X\Big)
\bigg]{V^2\over Z^{2m/(m-1)}}\bigg\}'=\cr
&\bigg[(m-2)(m-3)Z^4-2\Big((m^2-m+2)Y^2-(m-2)(m-3)X^2-4YW\cr
&-4(m-2)XW\Big)Z^2+\ter(X-Y)^2\Big(\,3(m^2+3m-2)Y^2-(m-2)(m-3)X^2\cr
&-2(m-2)(m-3)XY-8(2m-1)YW-8(m-2)XW\Big)\bigg]{V^2\over Z^{2m/(m-1)}}.&(4.8)}
$$

Of course for $m=2$ the equations reduce to those of the previous
section.
Again, (4.6) gives rise to a first integral:
$$\eqalignno{
&2Y-W+2\bigg\{\Big[(m^2-m+2)Y-2W\Big]Z^2+(X-Y)\Big[(m^2+3m-2)Y^2\cr
&-(m^2-m+2)XY+2XW-2(2m-1)YW\Big]\bigg\}{V^2\over Z^{2m/(m-1)}}=a.&(4.9)}
$$

The factor $V^2/Z^{2m/(m-1)}$ in the equations (4.6)-(4.8) can
be eliminated by use of (4.5). In this way one obtains rational
expressions in the variables $X$, $Y$, $W$ and $Z$. The equations
can then be cast in the canonical form
$$X'=F(X,Y,W,Z),\qquad Y'=G(X,Y,W,Z),\qquad W'=H(X,Y,W,Z).\eqno(4.10)$$
As before, the first integral (4.9) may be used to eliminate from the system
the variable $W$ and
reduce the problem to a three-dimensional dynamical system,
but we prefer to consider the phase
space as a three-dimensional surface embedded in four-dimensional space.
Also, one must be careful with the possible singularities
introduced by solving (4.5) with respect to $V^2/Z^{2m(m-1)}$.

The calculations were performed with MATHEMATICA. Unfortunately, the
expressions of the functions $F$, $G$ and $H$ contain hundreds of terms and
remain unmanageable also using computer algebra: our analysis is therefore
necessarily limited by this problem.

The critical points of the \dsy at finite distance are
\medskip
I) $Z_0=0$, $P_0=0$, and hence $V_0=0$. The constraint (4.9) then
implies $W_0=2Y_0-a$ and $X_0=\pm\sqrt{2Y_0^2-2aY_0+a^2}$. These
points correspond to a regular horizon only if $X_0=Y_0=W_0=a$.
\medskip
II) A generalization of the point II in $m=2$, which is defined by
$X_0=0$, $Z_0^2=2Y_0W_0-2Y_0^2-W_0^2$,
with $Y_0$ and $W_0$ related by the equation$\quad
(m-2)(m-3)W_0^4-4(m-1)(m-4)W_0^3Y_0+2(3m^2-19m+14)W_0^2Y_0^2$
$-{4\over3}(m^2-23m+16)W_0Y_0^3+(m^2-13m+14)Y_0^4=0.\quad$ As for $m=2$,
these points correspond to a finite value of $V_0$ only if
$Y_0=W_0=0$, where they reduce to a special case of I.
\bigskip

The phase space at infinity may be studied like for $m=2$ by defining the
new variables $u$, $y$, $w$ and $z$ and taking the limit $u\to 0$.
The field equations at infinity can then be written as
$$\dot u=-fu,\quad\dot y=g-fy,\quad\dot w=h-fw,\quad\dot z=(m-1-f)z,
\eqno(4.11)$$
with the notation introduced in the previous section. The critical
points correspond to the vanishing of the derivatives. This
happens for

I) $w_0=y_0=0$, $z_0=\pm1$, with $v_0$ undetermined.

II) $z_0=p_0=0$, and hence $y_0={1\over\sqrt2}$, $w_0=\sqrt2$, $v_0=0$.

{\noindent Due to the size of the equations involved, we cannot however
exclude the presence of further critical points, although this seems unlikely.}

For this reason, we do not discuss in detail the linearization of
the field equations near the critical points, although the results
look similar to the four-dimensional ones.
Also the \ab near the
critical points is analogous to $m=2$. In case I, $\x\to\x_0$, and
$$e^\z\sim|\x-\x_0|^{-1/(m-1)},\quad e^\n\sim \const,\quad e^\c\sim\const,$$
and hence, defining $r\sim|\x-\x_0|^{-1/(m-1)}$ and recalling (4.2),
$$ds^2\sim-dt^2+dr^2+r^2d\O_m^2,\qquad\f\sim\const.$$
At points II, $\x\to\pm\inf$, and
$$e^\z\sim e^{\mp\x},\quad e^\n\sim e^{\mp\x/\sqrt2},
\quad e^\c\sim e^{\mp\sqrt2\,\x},$$
and defining $r\sim e^{\mp[m+(m-2)/\sqrt2)]\x}$,
$$\eqalign {ds^2\sim&-r^{2(m-1)/[(m-2)+\sqrt2m]}dt^2+
r^{-2(m-1)/[(m-2)+\sqrt2m]}dr^2+
r^{-2(1-\sqrt2)/[(m-2)+\sqrt2m]}d\O_m^2,\cr
\qquad\ef\sim&\ r^{2/[(m-2)+\sqrt2m]}.}$$

We can conclude that regular \af \bh solutions exist also in $D>4$,
but we cannot completely exclude the presence of further solutions with
different asymptotics. Asymptotically (anti)-de Sitter solutions
can nevertheless be ruled out by explicit check.

\section{5. Conclusions}
We have studied the \dsy generated by the equations that determine
the \ssy solutions of EH-GB dilatonic theories and shown that in
four dimensions the only regular \bh  solutions are asymptotically
flat. In $D>4$ a complete answer was not possible due to the
complexity of the system, but it seems plausible that the same
property holds.

In spite of the difficulty of studying the case $D>4$,
we find however important the demonstration that methods from the
theory of dynamical systems can be used also in the case of
higher-derivative models of the Lovelock type.
In particular, these methods may be
used to study more involved problems, as for example black hole
solutions arising from the compactification of \hdim EH-GB models,
or cosmological models.

\section{Appendix}
In this appendix we give some details of a more rigorous treatment of phase
space in the four-dimensional case. This can be achieved if, instead of
confining to the physical hypersurface $E=0$, one considers the full
phase space.
In this way, the only singularity is at $Z=0$, and it is easier to control
the behaviour of the trajectories near the critical points.

Since we are not aware of any treatment of singular \dsy in the literature,
the only possibility to proceed rigorously would be to define new variables
such that the system becomes regular, for example introducing $\tV=V/Z^2$.
These variables however are not suitable for discussing the physics, since
$\tV$ would diverge in physically interesting situations, as for example at
the horizon of Schwarzschild-like solutions. For this reason we shall attempt
to maintain the old variables, taking the limit $Z\to0$ in the correct way,
when ambiguities may occur.

Our starting point are equations (2.4)-(2.8).
Since $E$ in (2.5) is a first integral of the dynamical system, all the
trajectories are confined to surfaces of constant $E$, but only those with
$E=0$ are physical. We shall therefore consider only the critical points lying
on this surface.
Equations (2.6)-(2.8) can be explicited for $X'$, $Y'$ and $W'$, as in (2.16):
$$X'=\bar F(X,Y,W,V,Z),\qquad Y'=\bar G(X,Y,W,V,Z),\qquad W'=\bar H(X,Y,W,V,Z),
\eqno(A.1)$$
where $\bar F$, $\bar G$ and $\bar H$ are rational functions of their variables.

Let us consider the critical points at finite distance first. These are placed
at $Z_0=V_0=0$. To simplify the discussion, we shall only consider the points
corresponding to regular horizons, where $X_0=Y_0=W_0=a$, and
$$Z\sim V\sim e^{a\x},\eqno(A.2)$$
with $\x\to\pm\inf$, depending on the sign of $a$, as explained in sect.\ 3.
In view of (A.2), we linearize the equations (2.4) and (2.6)-(2.8) around the
point $X_0=Y_0=W_0=a$, $V_0=kZ_0\to0$, for finite $k$, obtaining
$$\d Y'=\d X'=\d W'=4a\,\d Y,\quad\d Z'=a\,\d Z,\quad\d V'=a\,\d V,
\eqno(A.3)$$
with eigenvalues $0$ (2), $a$ (2), $4a$, independent of $k$. The zero
eigenvalues are due to the presence of two constants of motion. It results
that the critical point is  an attractor for $a<0$ and a repellor for $a>0$.

As discussed in sect.\ 3, the phase space at infinity can be studied by
defining new variables
$$u={1\over X},\quad y={Y\over X},\quad w={W\over X},\quad z={Z\over X},
\quad v={V\over X}.$$
The \fe at infinity ($u\to0$) read
$$\dot u=-fu,\qquad\dot y=g-fy,\qquad\dot w=h-fw,\qquad\dot z=(1-f)z,\qquad
\dot v=(w-f)v,$$
and the critical points are of course those obtained in section 3:

I) $y_0=w_0=0$, $z_0=\pm1$, any $v_0$, i.e.\ $X_0=\pm\inf$, $Y_0=W_0=0$,
$Z_0=\pm X_0$, any $V_0$.

II) $z_0=0$, $v_0=0$; for the physical solutions $y_0={1\over\sqrt2}$,
$w_0=\pm\sqrt2$. In terms of the original variables, $X_0=\pm\inf$,
$Y_0={X_0\over\sqrt2}$, $W_0=\sqrt2\,X_0$, $Z_0=V_0=0$.

The linearization around points I poses no problem, yielding
$$\d\dot u=-\d u,\quad\d\dot y=-\d y,\quad\d\dot w=-\d w,\quad\d\dot v=-\d v,
\quad\d\dot z=-2\,\d z.\eqno(A.4)$$
The eigenvalues are therefore all negative and for $X_0$=$\inf$ ($X_0$=$-\inf$)
these points attract (repel) the trajectories both from finite distance and from
infinity (remember that\ $\dot u\ =u\, d/d\x$).

At points II, for $\x\to\pm\inf$, $Z\sim e^{\mp\x}$, $V\sim e^{\mp\sqrt2\x}$, and
hence $v/z\to0$. Therefore, one must take the limit $v\to0$ first, and then
$z\to0$, getting
$$\d\dot u=\d\dot y=\d\dot w=0,\quad\d\dot v=\sqrt2\,\d v,\quad\d\dot z=\d z.
\eqno(A.5)$$
The eigenvalues are either null or positive. These points are therefore centers
for trajectories coming from finite distance, and repellors (actractors) for those
with $X_0$=$\inf$ ($X_0$=$-\inf$).

The resulting phase space portrait is rather similar to that of the EH-scalar
system, except for the location of the critical points II at infinity, that in
the EH case are placed at $y_0=w_0=0$. In particular, all the trajectories
corresponding to solutions with regular horizons start from the critical point
at $Z=V=0$, $X=Y=W=a>0$ for a parameter $a$, ending at $Z=X=\inf$, $Y=W=0$, and
hence are asymptotically flat. These solutions also possess $\x$-reversed copies
that start at $Z=X=-\inf$, $Y=W=0$, and terminate at $Z=V=0$, $X=Y=W=a<0$, with
identical properties.

\beginref
\ref [1] D. Lovelock, \JMP{12}, 498 (1971).
\ref [2] C. Lanczos, \AoM{39}, 842 (1938).
\ref [3] B. Zwiebach, \PL{B156}, 315 (1985);
B. Zumino, \PRep{137}, 109 (1986).
\ref [4] J. Madore, \PL{A110}, 289 (1985);
F. M\"uller-Hoissen, \PL{B163}, 106 (1985);
S. Mignemi, \MPL{A1}, 337 (1986).
\ref [5] D.J. Gross and J.H. Sloan, \NP{B291}, 41 (1987).
\ref [6] J. Madore, \PL{A111}, 283 (1985);
D. Bailin, A. Love and D. Wong, \PL{B165}, 270 (1985);
N.R. Stewart, \CQG{8}, 1701 (1991).
\ref [7] N. Deruelle and T. Dole\v zel, \PR{D62}, 103502 (2000);
C. Charmousis and J.F. Dufaux, \CQG{19}, 4671 (2002);
S.C. Davis, \PR{D67}, 024030 (2003);
J.P. Gregory and A. Padilla, \CQG{20}, 4221 (2003).
\ref [8] J.D. Bekenstein, \PR{D5}, 1239; 2403 (1972).
\ref [9] D.G. Boulware and S. Deser, \PRL{55}, 2656 (1985);
J.T. Wheeler, \NP{B268}, 737 (1986);
D.L. Wiltshire, \PL{B169}, 36 (1986).
\ref [10] S. Mignemi and N.R. Stewart, \PR{D47}, 5259 (1993).
\ref [11] C.G. Callan, R.C. Myers, M.J. Perry, \NP{B311}, 673 (1988).
\ref [12] P. Kanti,  N.E. Mavromatos, J. Rizos,  K. Tamvakis and
E. Winstanley, \PR{D54}, 5049 (1996);
T.  Torii,  H.  Yajima, and K. Maeda, \PR{D55}, 739 (1997);
S.O. Alexeyev and M.V. Pomazanov, \PR{D55}, 2110 (1997).
\ref [13] S. Alexeyev and S. Mignemi, \CQG{18}, 4165 (2001).
\ref [14] D.G. Boulware and S. Deser, \PL{B175}, 409 (1986).
\ref [15] S. Mignemi and D.L. Wiltshire, \CQG{6}, 987 (1989);
D.L. Wiltshire, \PR{D44}, 1100 (1991);
S. Mignemi and D.L. Wiltshire, \PR{D46}, 1475 (1992);
S.J. Poletti and D.L. Wiltshire, \PR{D50}, 7260 (1994);
S. Mignemi, \PR{D62}, 024014 (2000).
\endref

\end